\documentstyle[12pt]{article}
\def\beq{\begin{equation}}
\def\eeq{\end{equation}}
\def\bea{\begin{eqnarray}}
\def\eea{\end{eqnarray}}

\begin{document}

\vspace{1cm}
\title{
%\vspace{4cm}
%\begin{flushleft}
{\Large \bf Multidimensional Einstein-Yang-Mills cosmological models
\thanks{Talk presented by Y. K. at Trobades Cient\'{\i}fiques de la
Mediterr\`ania. Encuentros Relativistas Espa\~noles E.R.E. 94, Ma\'o,
Menorca, Spain, 12-14 September 1994.}}} %\end{flushleft} }
\vspace{17mm} \author{Yuri Kubyshin \thanks{On leave of absence
from Nuclear Physics Institute, Moscow State University, 119899 Moscow,
Russia.} \thanks{E-mail address: kubyshin@Ecm.Ub.Es},
Enrique Moreno \thanks{E-mail address: moreno@Ecm.Ub.Es} \\
{\em Departament d'Estructura i Constituents de la Materia}\\
{\em Universitat de Barcelona}\\ {\em Av. Diagonal 647,
08028 Barcelona, Spain} \\
and \\
Jos\'e Ignacio P\'erez Cadenas \\
{\em Nuclear Physics Institute, Moscow State University}\\
{\em Moscow 119899, Russia} }
\date{October 11, 1994}
\vspace{0.7cm}
\maketitle
\setlength{\baselineskip}{2.6ex}
\begin{abstract}
We study the process of the evolution of the space of extra dimensions
in the framework
of Einstein-Yang-Mills cosmological models. It is shown that, for certain
classes of models, the static compact space of extra dimensions is the
attractor for a wide range of initial conditions. Also the effect of
isotropization of extra dimensions in the course of evolution is
demonstrated.
\end{abstract}

\vspace{1cm}

{\bf 1.} The Kaluza-Klein theories, i.e. theories formulated on the
space-time with $(1+3+d)$ dimensions, is an important framework for
the unification of particle interactions \cite{KK} and,
probably, could be relevant in cosmology for
generation of anisotropic and inhomogeneous generalizations of the
radiative Friedmann - Lema\^{\i}tre - Robertson - Walker (FLRW)
models in four dimensions \cite{verdaguer}. When considered in the
cosmological setting a Kaluza-Klein
model should possess certain classical properties in order not to
contradict the observations. The most important of them are the following:

a) The model should possess a solution describing an expanding three-
dimensional space to the present size of the Universe.

b) The scale factor(s) $L$ of the space of extra dimensions
should decrease up to a size $L \leq (0.1 \div 1) TeV^{-1}$
in the course of the evolution, described by the same solution,
and the model should have a mechanism
for stopping the contraction or at least slowing it down considerably at
this value of $L$.
The bounds on the variation of $L(t)$ following from the yield
of primordial $^{4}He$ are: $0.99 \leq L_{0}/L_{N} \leq 1.01$,
where $L_{0}$ is the present size of the space of extra
dimensions and $L_{N}$ its size at the time of the nucleosynthesis
\cite{kolb-book}.

c) The solution should be an attractor for a wide
range of initial conditions (naturallness of the solution).

d) The solution
should satisfy restrictions imposed by the limits on the energy
density of particles produced during the contraction of the
space of extra dimensions (see Refs. \cite{p-prod}).

We would like to remark that many of the multidimensional
cosmological models considered in the literature do not exhibit
some and in some cases none of these basic properties.
Examples of these are
Kasner-type solutions in $(4+d)$ dimensions and models
dominated by the energy
density of radiation (see eg. \cite{kolb-book}).

{\bf 2.} In this contribution we consider a class of Einstein-Yang-Mills
models which will be shown to satisfy the conditions (b)-(c). The
isotropization of the space of extra dimensions metric
will be also shown.

The models are formulated on the manifold $M_{(4+d)} = M_{(4)}
\times K$, where $M_{(4)}$ is the four-dimensional part of the
space-time and $K=S/R$ is a compact homogeneous $d$-dimensional space
on which the group $S$ acts transitively. The action is
given by
\begin{equation}
{\cal S} = \int_{M_{(4+d)}} d \hat{x}
      \sqrt{-\hat{g}} \left( \frac{\hat{R}}{16 \pi \hat{\kappa}}
     - \frac{1}{8 \hat{e}^{2}} Tr \hat{F}_{MN} \hat{F}^{MN}
    - \frac{\hat{\Lambda}}{16 \pi \hat{\kappa}}\right), \label{action1}
\end{equation}
where $M,N = 0,1,\ldots 3+d$, $\hat{g} = \det (\hat{g}_{MN})$, $\hat{R}$
is the multidimensional scalar curvature,
$\hat{\kappa}$, $\hat{\Lambda}$ and $\hat{e}$ are the gravitational,
cosmological and gauge constants in $(4+d)$ dimensions respectively and
$\hat{F}_{MN}$ is the stress tensor of the gauge field with a gauge
group $G$. We consider the class of models with metric
$\hat{g}_{MN}$ given by:
\begin{equation}
ds^2 = - dt^2 + a(t)^2 d \Omega_{(3)}^2 + L_{1}^2(t)
       d \omega_{(d_{1})}^2
       + \ldots + L_{k}^2(t) d\omega_{(d_{k})}^2.
       \label{metric}
\end{equation}
The four-dimensional part of the metric is the one corresponding to a
FLRW model and
the metric on $S/R$ is the most general $S$-invariant metric.
The terms $L_{p}^2 d \omega_{(d_{p})}^2$ determine
the invariant metric on the irreducuble
invariant $d_{p}$-dimensional subspaces in the
tangent space decomposition of $S/R$ under the action of the
isotropy group $R$.

We assume that there is a non-trivial background gauge field,
which is described by the so-called $S$-symmetric gauge field
configuration, satisfying the equations of motion.
An important solution for the Kaluza-Klein
cosmology of this type, is the one corresponding to magnetic
monopole like configurations on $S/R$,
and this was first proposed to drive the compactification of the extra
dimensions in \cite{cremmer}. The dimensional reduction of the theory
(\ref{action1}) can be carried out consistently and gives
(see \cite{KPC}) an effective theory in four
dimensions describing Einstein gravity with scalar fields
associated to the scales $L_{p}(t)$ with a certain potential
$W$. For the models considered here the potential posseses at least
one minimum due to the non-trivial monopole contribution.
In general, depending on the initial conditions,
the evolution leads to one of the two situations:

i) All or some of the scales $L_{p}$ tend to $+\infty$. This
corresponds to decompactification, i.e. unlimited growth of the
size of extra dimensions, and is physically unacceptable.

ii) The scales approach the minimum, $L_{p} \rightarrow L_{p0}$,
through damped oscillations about it (see below). This corresponds
to a successful compactification of the extra dimensions.

The question we address here is how large is the region
of initial values
$(L_{p}(0)$, $\dot{L}_{p}(0))$ for which compactification is
successful, or in other words, for which set of initial conditions
the minimum is an attractor.

{\bf 3.} We consider first the case where $S/R$ is a symmetric space
\cite{KRT,BKM}.
Then the metric is parametrized by one single scale, the scalar curvature
on $S/R$ is $R^{(d)}={\cal R}_{0}/L^2$ and the monopole
contribution is given by $Tr F^2/(8 e^2) = v_{0}/L^4$, where
${\cal R}_{0}$ and $v_{0}$ are some constants.

For the range of values of $\hat{\Lambda}$ that we are of interest
here the potential
has a minimum at $L=L_{min}$ and a maximum at $L=L_{max}$.
We introduce the scalar field $b(t) = \ln (L(t)/L_{min})$ and
choose the cosmological constant such that $W_{min} \equiv
W(b=0) = 0$. Then $L_{min} = 2 v_{0}/{\cal R}_{0}$, $L_{max}^2 =
(d+4) L_{min}^2/d$ and
the potential is equal to
\begin{equation}
 W(b) = \frac{{\cal A}}{16 \pi \kappa} e^{-db}
            \left(1 - e^{-2b} \right)^{2},
                              \label{wsymm}
\end{equation}
where ${\cal A} = {\cal R}_{0}^2/(4 v_{0})$.
The equations of motion are
\begin{eqnarray}
  h^{2} & \equiv & \left(\frac{a'}{a}\right)^2 =
          \frac{d(d+2)}{12} b'^2 + (16 \pi \kappa)
	  \frac{W(b)}{6},   \label{eqsymm1} \\
  & & b'' + 3 h(t) b' + \frac{16 \pi \kappa}
  {d(d+2)} \frac{d W(b)}{d b} = 0,   \label{eqsymm2}
\end{eqnarray}
where the prime means derivative with respect to $t$. This system
was studied qualitatively in refs. \cite{KRT,BKM}.
It describes the evolution of a homogeneous scalar field
in the potential (\ref{wsymm}) with a growing scale factor $a(t)$.
The quantity
\begin{equation}
   E=h^2/M_{Pl}^4= d(d+2) b'^2 / 2 M_{Pl}^2 +
   W(b) / M_{Pl}^4             \label{esymm}
\end{equation}
can be interpreted as the energy of a 1-dimensional particle
(in the units of $M_{Pl}$).
The motion of the particle is dissipative due to the viscosity term
in eq. (\ref{eqsymm2}). Indeed, that equation can be rewritten as
$\dot{E} = - 3 h b'^2 \leq 0$.

In Fig. 1 we show part of
the phase plane $(b, b')$ around the
minimum for $d=6$. The dashed lines represent the levels of constant
$E$: $E=1$ and $E=10$. The former one corresponds
to the Planckian energy densities and thus limits the sector of the
classical dynamics, i.e. the region of values of $b$ and $b'$
where the energy density of the field is smaller than $M_{Pl}^4$ and
hence the classical equations of motion make sense.
We suppose that the Universe started its classical evolution from
small values of $L$ and we consider the case where the initial value
$b(0) = b_{0} < 0$. Fig. 1 shows two typical
trajectories (thin lines) corresponding to the
cases (i) and (ii) of Sect. 2.

The thick line in Fig. 1 is the boundary of the
trapped region $T$, i.e. the part of
the phase space for which the minimum $b=0$ is the attractor.
Lacking a satisfactory quantum theory of gravity it is
not possible to know the intial conditions for the classical
evolution of the Universe. If we follow the conventional
Hot Big Bang scenario, the trapped region in Fig. 1
corresponds to the region where the initial conditions at the moment,
when the Universe enters the low temperature regime,
should lie for the compactification to be successful.
We see that the large part of the classical region lies inside $T$.

Consider now scenarios based on the quantum creation
of the Universe. In this case, according to the
Vilenkin distribution for $b$ \cite{vilenkin} or the
Hartle-Hawking distribution function \cite{hartle-hawking},
the Universe was probably created
with $b'=0$ away from the origin of the potential, i.e.
where $E = W(b)/M_{Pl}^4 \sim 1$, or on the top
of the local maximum of the potential.
(These statements, however, should be taken with caution
since the potential (\ref{wsymm}) strictly speaking
does not satisfy the condition of slow variation.)
Fig. 1 shows that in both cases the initial state is within the
trapped region and the Universe always approaches the
$M^{(4)} \times S/R$ space-time with $L=L_{min}$. For other
numbers of the extra dimensions the situation is qualitatively
the same, but for increasing $d$  the
region $T$ extends itself towards the direction of negative values for $b$.
Similar analysis for the Candelas - Weinberg
model and the Calabi-Yau compactification in a superstring model
can be found in \cite{maeda-pang}.

{\bf 4.} $S/R = SU(5)/SU(2) \times U(1)$.
The space of extra dimensions is a six-dimensional
non-symmetric homogeneous space, known as the
$CP^3$ manifold, with the
anisotropic metric (\ref{metric}) parametrized by
two scales ($k=2$).
After fine tuning $\hat{\Lambda}$, as in Sect. 3, the
potential, $W$, has a minimum at $(L_{1min}, L_{2min})$ with
$L_{2min}^2=2L_{1min}^2$ and a saddle point with $L_{2s}^2
= 2 L_{1s}^2=10 L_{1min}^2/3$. Its explicit form
as a function of the two scalar fields,
$b_{p}(t) = \ln L_{p}(t)/L_{pmin}$ ($p=1,2$),
was calculated in \cite{KPC}. Fig. 2 represents the contour plot of
the potential in
the $(b_{1},b_{2})$-plane and the levels $E=1$ and $E=10$,
where, $E \equiv W(b_{1},b_{2})/M_{Pl}^4$.
$W$ decreases exponentially for
$b_{1},b_{2} \rightarrow +\infty$ and grows exponentially
when $b_{1},b_{2} \rightarrow -\infty$. The line $b_{1}=
b_{2}$ corresponds to the isotropic
$SU(4)$-invariant metric on $SU(4)/SU(3) \times U(1)$
(unsquashed $CP^3$), which is of course topologically
equivalent to $S/R$. The thick closed curve is the boundary of
the trapped region, which in
this case, is the region of initial conditions
$(b_{10},b'_{10}=0;b_{20},b'_{20}=0)$ for
which compactification is successful. For {\cal all}
initial conditions of this type, with $b_{10},b_{20} < 0$
situated in the classical region
$E \leq 1$, the minimum of the potential $b_{1}=b_{2}=0$
is the attractor. The same holds for initial conditions
favoured by scenarios
of the quantum creation of the Universe.
It appeares that, in this model,
isotropy in the space of extra dimensions is restored dynamically.
Two typical trajectories with highly anisotropic initial metric
(i.e. with $b_{1} \neq b_{2}$) are depicted in Fig. 2. We see
that after a few oscillations the trajectories approach the line
of isotropic metrics, for which $b_{1}=b_{2}$, and the main
part of evolution occurs in the narrow vicinity of this line.

{\bf 5.} Thus we have shown that the Einstein-Yang-Mills cosmological
models considered here satisfy conditions (b) and (c) of
Sect 1. Indeed, a $M^4 \times S/R$ Universe with constant
radii for the space of extra dimensions
is achieved for a wide range of initial conditions in the
Hot Big Bang scenario and for the initial states
suggested by scenarios of quantum creation of the Universe. Furthermore,
in the model with two
scales in the space $K$, it was shown that
isotropization occurs, i.e. the metric
tends to the one with higher symmetry in the course
of its evolution.

What about conditions (a) and (d) above?
During the evolution of $L_{p}(t)$, the scale factor of the
3-dimensional space expands according to the law $a(t) \sim t^{\lambda}$
with $\lambda < 1$ \cite{KRT}. Hence, our models do not
describe inflationary expansion without,
for instance, additional scalar fields \cite{BKM,BBS}.
The inclusion of effects due to particle production (see \cite{maeda})
and vacuum polarization might also help in achieving
power law inflation in these models.  Once the scales
of the extra dimensions have reached their minimum, then the main
contribution to expansion comes from radiation and further evolution
of $a(t)$ follows the standard radiation-dominated scenarios.
To check whether
the models satisfy condition (d) further calculations are
needed. This work is in progress now.

\vspace{10mm}

\noindent{\large \bf Acknowledgements}

We would like to thank Orfeu Bertolami, Jaume Garriga, Jose Mour\~ao,
Jose M.M. Senovilla and Enric Verdaguer for valuable discussions
and comments. This work was supported by funds provided by
M.E.C (Spain) and by CIRIT (Generalitat de Catalunya).

\newpage

\section*{Figure captions}
\begin{description}

  \item[Fig. 1] Phase plane $(b',b)$ of the model with
   symmetric space of extra dimensions. The dashed
   curves represent lines of constant "energy" defined
   by eq. (\ref{esymm}). Thick solid line is the
   boundary of the trapped region $T$, where lie the
   points for which the Universe $M^4 \times S/R$,
   with $L=L_{min}$, is an attractor. The two characteristic
   trajectories are shown as thin solid lines.

   \item[Fig. 2] Contour plot of the potential, $W(b_{1},b_{2})$,
   of the model of Sect. 4. Curves of constant
   $E = W/M_{Pl}^4$ are shown as dashed lines. $E=1$
   marks the limit of application of the classical
   description. The thick closed curve is the
   boundary of the trapped region, the region for which
   the minimum of the potential is an attractor.
   Two characteristic trajectories of the evolution
   of the scale factors are shown as thin solid lines. The
   initial conditions are anisotropic $(b_{1} \neq
   b_{2})$, however the metric isotropizes after
   a few oscillations.

\end{description}

\end{document}